\def\cm2{cm$^{-2}$}

\def\c2{C~{\sc ii}}
\def\c4{C~{\sc iv}}
\def\fe2{Fe~{\sc ii}}
\def\fe3{Fe~{\sc iii}}
\def\mg1{Mg~{\sc i}}
\def\mg2{Mg~{\sc ii}}
\def\si2{Si~{\sc ii}}
\def\si4{Si~{\sc iv}}
\def\al2{Al~{\sc ii}}
\def\al3{Al~{\sc iii}}
\def\o1{O~{\sc i}}
\def\n1{N~{\sc i}}
\def\h1{H~{\sc i}}

\def\approxlt{\mathrel{\spose{\lower 3pt\hbox{$\sim$}}
        \raise 2.0pt\hbox{$<$}}}
\def\approxgt{\mathrel{\spose{\lower 3pt\hbox{$\sim$}}
        \raise 2.0pt\hbox{$>$}}}

\documentclass[article]{gwp80}
\usepackage{graphicx}
\usepackage{amssymb}
\tabletypesize{\normalsize}  % AMcW

\def\plotone#1{\centering \leavevmode
\includegraphics[width=.95\columnwidth]{#1}}

\def\plotone#1{\centering \leavevmode
\includegraphics[width=.95\columnwidth]{#1}}

\shortauthors{Stellingwerf}
\shorttitle{Period Determination of RR Lyrae Stars}
\begin{document}
\large
\pagenumbering{arabic}
\setcounter{page}{47}

\title{Period Determination of RR Lyrae Stars}

\author{{\noindent Robert F. Stellingwerf{$^{\rm 1,3}$}\\
\\
{\it (1) Stellingwerf Consulting, Huntsville, Alabama, USA} 
}
}

\email{(1) rfs@stellingwerf.com}

\begin{abstract}
The classic problem of detection of periodic signals in the presence of noise becomes much more challenging if the observation times are themselves periodic, contain large gaps, or consist of data from several different instruments. For RR Lyrae light curves the additional attribute of highly non-sinusoidal variation adds another dimension. This memo discusses and contrasts Discrete Fourier, Periodogram, and Phase Dispersion Minimization (PDM) analysis techniques. A new version of the PDM technique is described with tests and applications.
\end{abstract}

\section{Introduction}

This memo discusses the problem of detection of periodic components in variable star data sets. This type of data tends to be highly non-sinusoidal, the period and the amplitude may change secularly or periodically, there may be multiple modes present, other types of variations due to companions, eclipses, etc. could occur, and the data set usually contains gaps, which themselves may contain periodic variations (daily, yearly, etc.). Although we will focus on RR Lyrae data, the techniques are applicable to other classes of stars and, indeed, other application areas as well. 

\section{Types of Analyses}

\subsection{Fourier Transform (FT, FFT)}

Most analyses of periodic signals depend on the classical techniques of Discrete Fourier Transforms (FT). An FT analysis decomposes the time data into a sum of sine and/or cosine components whose coefficients represent the amplitude of each frequency in the expansion. These coefficients are computed as follows.

%[EQUATION 1]
{\Large
\begin{equation}
X_k = \sum_{n=0}^{N-1}  x_n e^{-\frac{2\pi i}{N}\, kn}    \hskip2.0cm          k= 0,....,N-1  
\end{equation}
}

%\begin{figure*}[h]
%\centering
%\includegraphics[width=12cm]{fig1.jpg}
%\plotone{EQ1.eps}
%\vskip0pt
%\end{figure*}

The squares of these amplitudes then describe a ``power spectrum'', i.e. power versus frequency,
function for the data set. See Press et al. (1992; henceforth NRC92), chapter 13, for details. For
long sets of equally spaced data this
approach is optimum. In real data sets, however, the time intervals are generally not equal, and
may contain large gaps due to observational constraints. This has led to the use of other techniques
applicable to non-equal-spaced data.

\subsection{Periodogram (PG)}

The ``periodogram'' is a classical analysis technique that employs an FT type of expression, but
ignores the ``equal spacing'' requirement. The power spectrum of the PG approach is computed thus:

%[EQUATION 2]

{\Large
\begin{equation}
P_x(\omega) = \frac{1}{N_0} \left[ \left( \sum_{j}  x_j\, {\rm cos}\, \omega t_j \right)^2 +
\left( \sum_{j}  x_j\, {\rm sin}\, \omega t_j \right)^2 \right] \\
\end{equation}
}

%\begin{figure*}[h]
%\centering
%\includegraphics[width=12cm]{fig1.jpg}
%\plotone{EQ2.eps}
%\vskip0pt
%\end{figure*}

Here $\{t_j x_j\}$ is the set of time and data values, $w$ is the frequency, and $N_0$ the number of data points. This technique has been thoroughly analyzed by Scargle (1982). 

A variation has been derived by Vanicek (1971) and Lomb (1976) with desirable statistical properties. This power spectrum is given by 

%[EQUATION 3]

{\Large
\begin{equation}
P_x(\omega) = \frac{1}{2}  \left( \frac{ \left[ \sum_{j}  x_j \,{\rm cos}\, \omega (t_j-\tau) \right]^2}
                             { \sum_j {\rm cos^2}\, \omega (t_j-\tau) }       
 +                                \frac{ \left[ \sum_{j}  x_j\, {\rm sin}\, \omega (t_j-\tau) \right]^2}
                             { \sum_j {\rm sin^2}\, \omega (t_j-\tau) }       
                          \right)
\end{equation}
}

%\begin{figure*}[h]
%\centering
%\includegraphics[width=12cm]{fig1.jpg}
%\plotone{EQ3.eps}
%\vskip0pt
%\end{figure*}

with

% [EQUATION 4].
{\Large
\begin{equation}
{\rm tan}\,2\omega\tau = \frac{ \sum_j {\rm sin}\, 2\omega t_j }
                              { \sum_j {\rm cos}\, 2\omega t_j }
\end{equation}
}

%\begin{figure*}
%\centering
%\includegraphics[width=12cm]{fig1.jpg}
%\plotone{EQ4.eps}
%\vskip0pt
%\end{figure*}

With the Lomb normalization the statistical distribution of P$_x$ for a pure Gaussian noise signal
is exponential. This is a useful property when estimating the ``significance'', or chance that
a detected signal is due to noise fluctuations. Lomb also showed that the periodogram was equivalent
to a least squares fit to the folded data at each frequency by a sine wave. This is illustrated
in Figure 1, below.

\begin{figure*}
\centering
\plotone{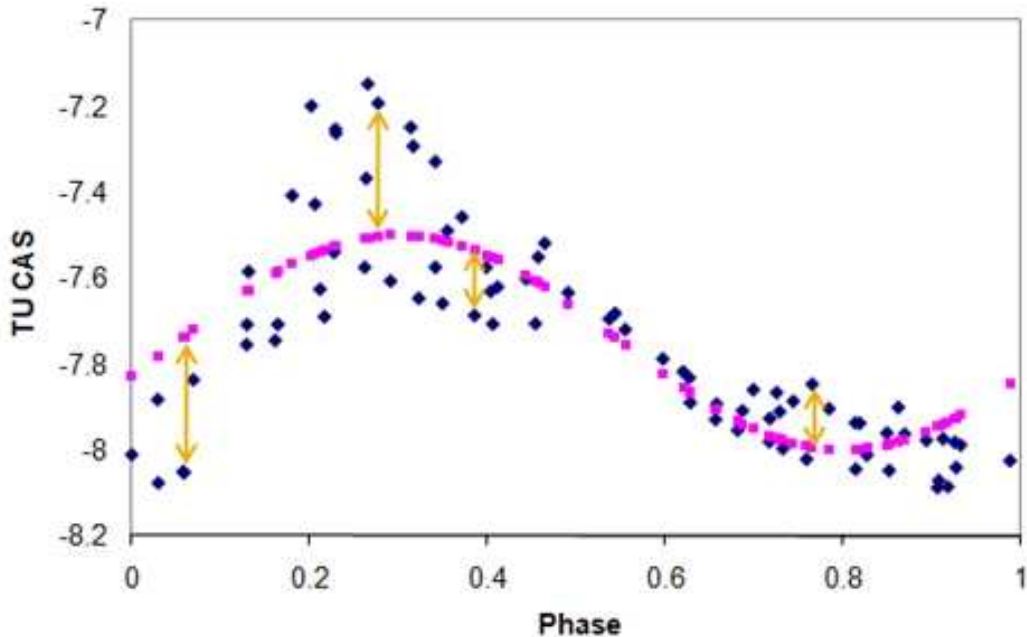}
\vskip0pt
\caption{Illustration of a periodogram analysis of folded light curve data. Arrows represent deviations from a sine wave fit.  }
\label{fig1}
\end{figure*}

\subsection{Phase Dispersion Minimization (PDM, PDM2)}

The ``Phase Dispersion Minimization'' technique was originally presented in Stellingwerf (1978).
This is basically a folding of the data, together with a binning analysis of the variance at
each candidate frequency. It is a least squares fit, as in the periodogram, but to a mean curve
that is determined by the data, rather than a sine wave.

\begin{figure*}
\centering
\plotone{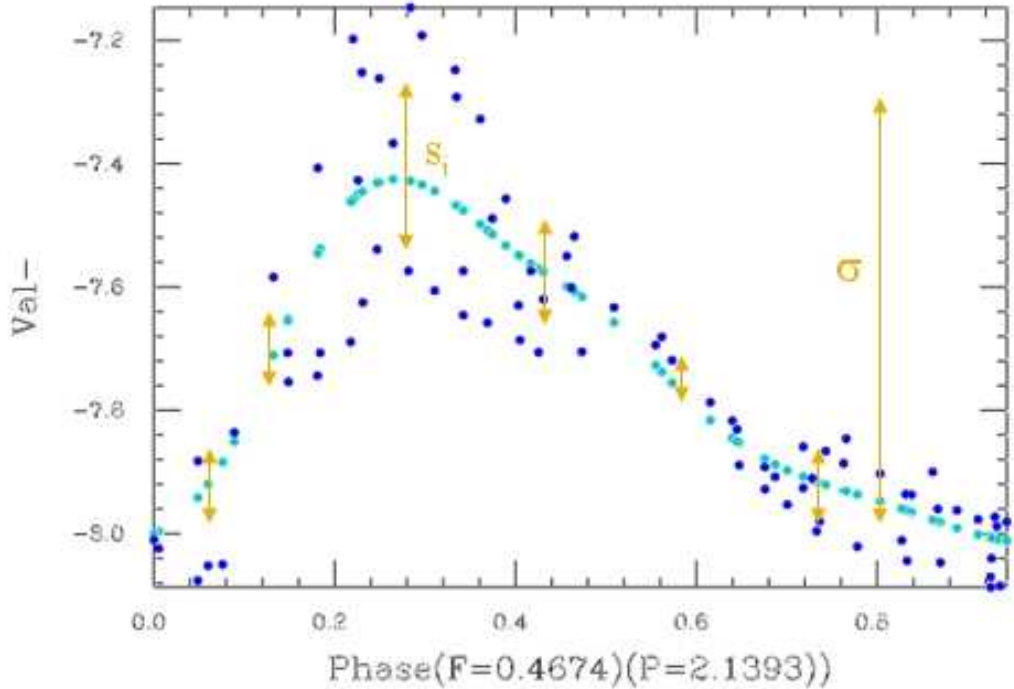}
\vskip0pt
\caption{Illustration of the PDM analysis of folded light curve data. Arrows represent the bin standard deviations (s.d.) and the full data s.d. }
\label{fig2}
\end{figure*}

Rather than a power curve, the PDM analysis computes the sum of the bin variances divided by the total variance of the data. For uncorrelated data this ratio is close to unity. At a possible period the bin variances are less than the overall variance, and the statistic (called Theta) drops to some value greater than zero. 

The PDM computation can be done very efficiently. For a time/data set $\{t_j x_j\}$ and candidate
frequency $f_i$, the phase $\phi_{ij}$ of a data point is the fractional part of $f_i * t_j$.  Assuming
10 bins, the bin number of this point is just the integer part of 10 * $\phi_{ij}$.  Once the bin number
is known, the statistics for this bin can be updated. Thus the computation consists of a single loop
over data points for each frequency. This is about 10 times more efficient than the original PDM code,
which contained nested loops over data and bins for maximum generality. Also note that no trigonometric 
function evaluations are required.

For data sets with large gaps in time, the PDM technique is usually applied to each cluster of points
(called a {\bf``segment''}) separately, greatly reducing the computation required, since the spectral
line width (which determines the number of frequencies needed) will be the time length of the longest
segment, rather than the total time span for the data set. Similarly, with appropriate adjustments in
scaling and zero point, different color observations can be combined in the same fashion. This is
illustrated in Figure 3.

\begin{figure*}
\centering
\plotone{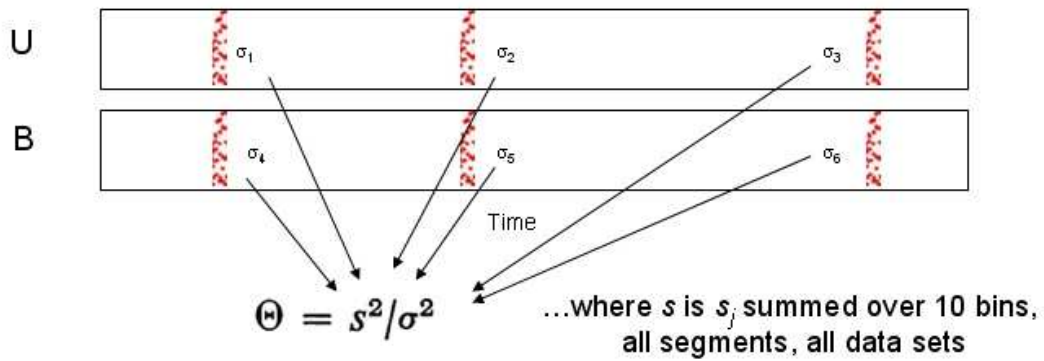}
\vskip0pt
\caption{
Illustration of a PDM analysis of a data set with multiple colors and large gaps.}
%Folded test data (top), Lomb power spectrum (center), and PDM Theta function (bottom). }
%Illustration of a PDM analysis of a data set with multiple colors and large gaps. }
\label{fig3}
\end{figure*}

Additional improvements to the original PDM technique have been made. The updated version, 
called {\bf PDM2}, is available at the website www.stellingwerf.com. The other improvements
and changes include: 

1.	Experience has shown that {\bf 10 bins} are adequate (and usually optimum) for variable star data sets. For data sets with more than about 100 points, these bins are non-overlapping. For data sets with less than about 100 points, better results are obtained if the bins are double-wide, and overlap. The switch point is adjustable, and one or the other approach can be specified if desired. One approach is to use overlapping bins to identify possible candidate frequencies, and then use the non-overlapping option to improve the period estimate and obtain a mean light curve.
 
2.	The statistical significance of each frequency is now computed either with a 
{\bf Beta distribution} (Schwarzenberg-Czerny, 1997) or a {\bf Monte Carlo} computation 
(Nemec \& Nemec 1985) rather than the original F test. This will be discussed below.

3.	The sum of the bin variance technique is equivalent to a least squares fit to a step function
through each bin mean. For clean data the results can be improved by fitting to a {\bf linear variation}
through the bin means, or by a {\bf B-spline curve} through the bin means. These are now options.
They are only used at frequencies where a significant signal is present.

4.	{\bf Subharmonic Averaging:} Unlike the Fourier techniques, PDM detects a signal at 1/2,
1/3, etc the actual frequency, since these multiple period folding frequencies also produce
periodic variations. This option looks for a significant minimum in Theta at both the main and
1/2 the main frequency. For actual variations, both will be present. For a noise result, the 1/2
frequency signal will not be present. The Theta variation for a sine wave with frequency 1.0
(overlapping bins) is shown in Figure 4, illustrating this effect.

5.	{\bf Period Variation:} PDM2 has an option to include a slow increase or decrease in the 
period across the data set. The rate of variation can be varied, and the most likely value for the
period change can be computed.

\begin{figure*}
\centering
\plotone{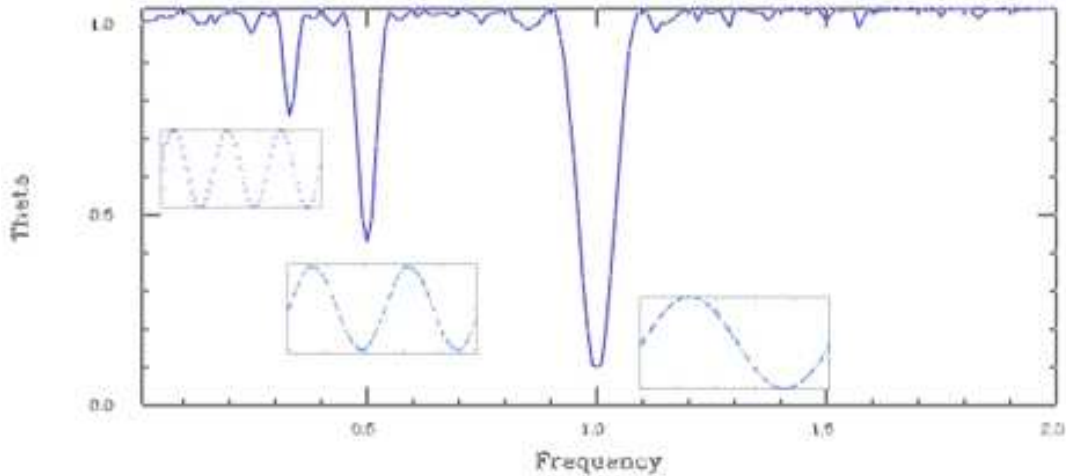}
\vskip0pt
\caption{PDM Theta function variation for a sine wave at frequency = 1. Inserts show the wave forms corresponding to the three main minima. }
\label{fig4}
\end{figure*}

\section{Comparison of PDM and Lomb Techniques}

Here we compare the Lomb periodogram and the PDM techniques for various test cases. The first data set consists of 10 cycles of a sine wave plus Gaussian noise. Figure 5 shows the folded data, the Lomb power spectrum, and the PDM Theta function for this case. Aside from the flip and the PDM subharmonics, the variation is almost identical in the two cases. We conclude that either approach will work for a simple signal plus noise case.

\begin{figure*}
\centering
\plotone{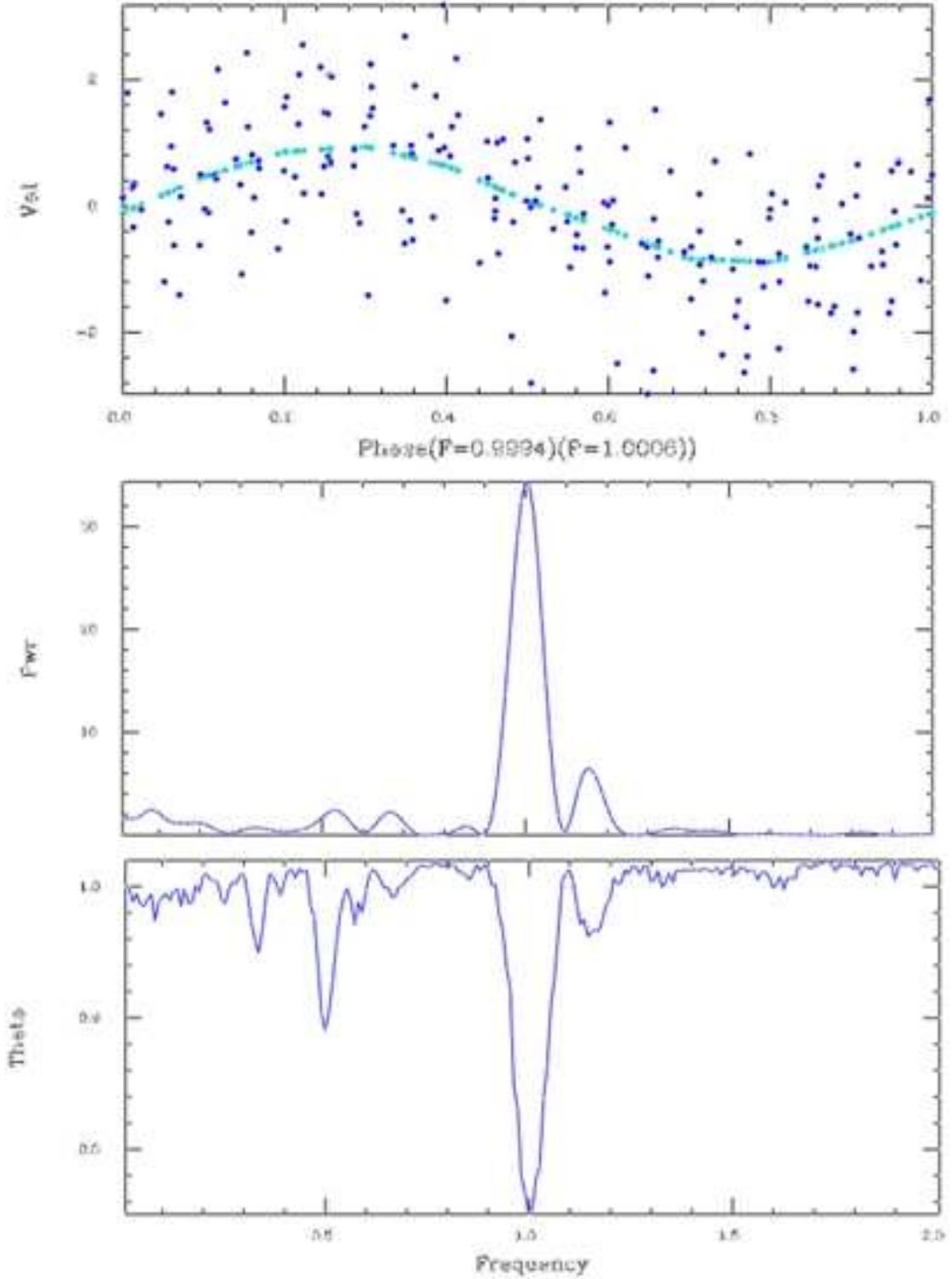}
\vskip0pt
\caption{
Folded test data (top), Lomb power spectrum (center), and PDM Theta function (bottom).}
%Illustration of a PDM analysis of a data set with multiple colors and large gaps. }
\label{fig5}
\end{figure*}

The effect of a large gap in the data is shown in Figure 6. Here the data consists of 10 cycles of a frequency = 1 variation, followed by a gap of about 50, then 10 additional cycles. No noise was added. Again, the Lomb and the PDM responses are very similar. Subharmonic averaging was used in this PDM analysis, which emphasizes the correct minimum over the adjacent frequencies.

Figure 7 shows a PDM analysis of the same data set, but with the data treated as three segments. This is the default mode of analysis, and the same result is obtained much more easily since the fine structure is eliminated. Normally this analysis would be followed by a pass without segmentation, but covering only the central dip seen in Figure 6 to improve the period estimate.

\begin{figure*}
\centering
\plotone{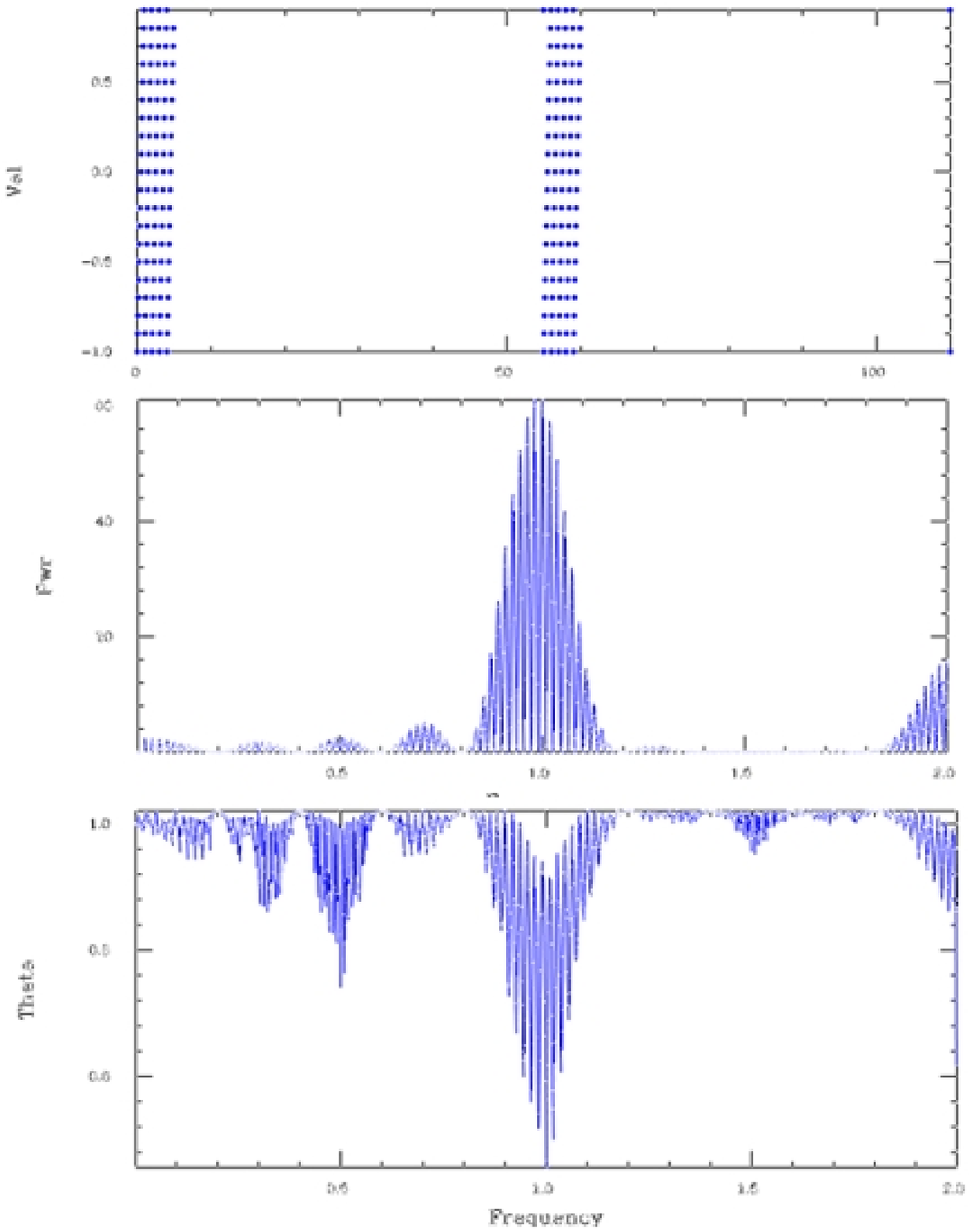}
\vskip0pt
\caption{Data set with large gap (top), Lomb power spectrum (center), and PDM response (bottom). }
\label{fig6}
\end{figure*}

\begin{figure*}
\centering
\plotone{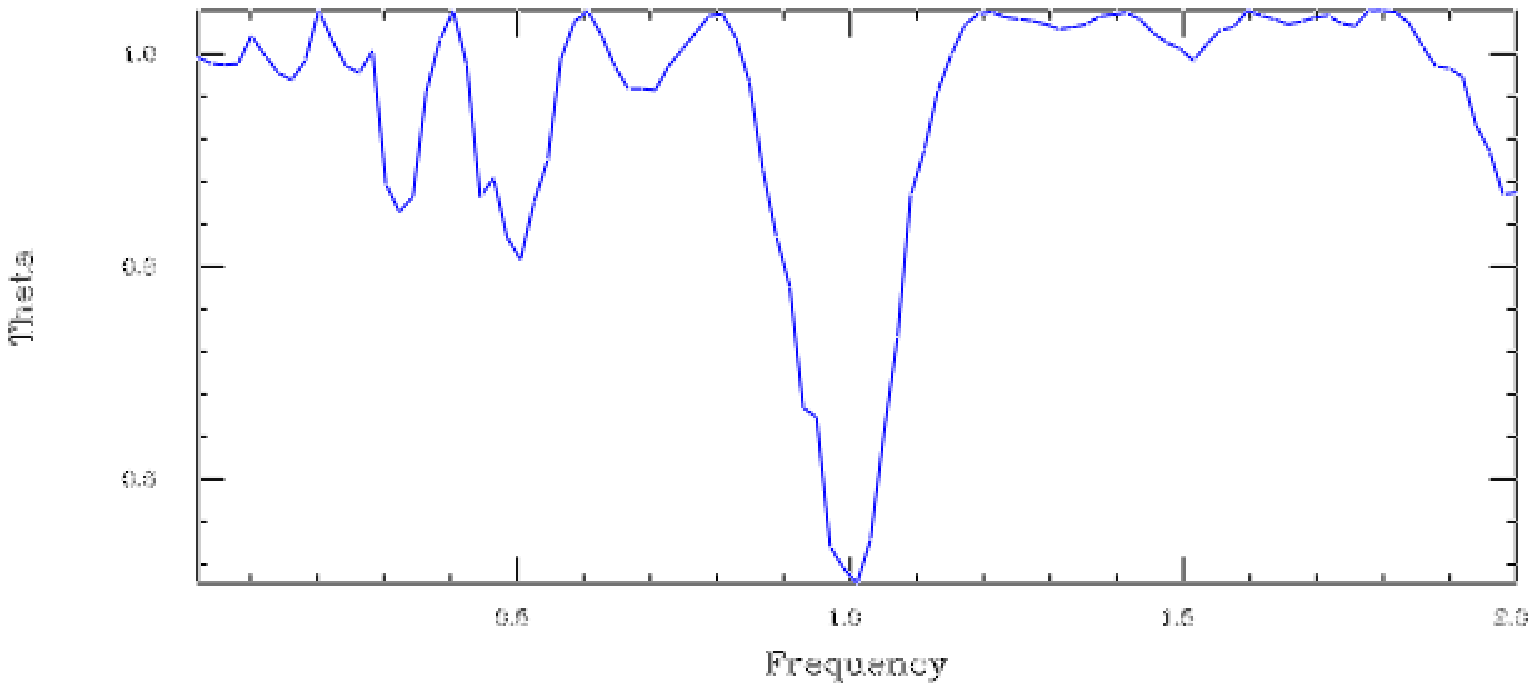}
\vskip0pt
\caption{ PDM analysis of the same data set as Figure 6, but with data segmentation option in effect.}
\label{fig7}
\end{figure*}

The next test consists of a series of narrow pulses (short duration eclipses would be similar). Figure 8 shows that the two techniques each get the correct answer, but the curves do look quite different for this case. Here the PDM result is much more definitive, with narrower spectral lines and smaller side lobes.

\begin{figure*}
\centering
\plotone{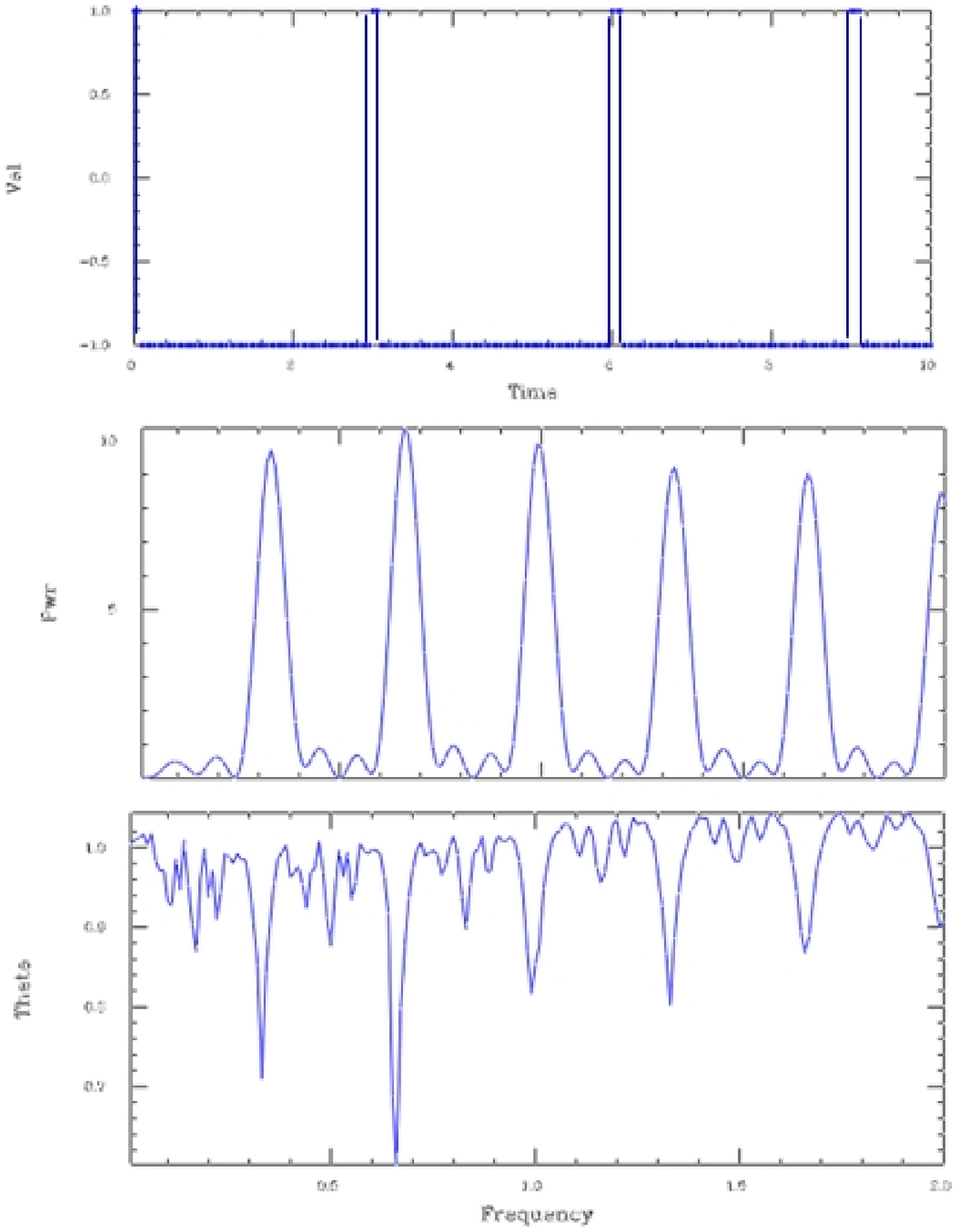}
\vskip0pt
\caption{ Analysis of a pulse data set: data (top), Lomb power spectrum (center), PDM Theta function (bottom).
}
\label{fig8}
\end{figure*}

A final test case shows that for complex wave forms, divergent results can be obtained. This
case (Figure 9) has alternating peaks, so that the periodic signal is at period = 2. The Lomb
technique identifies most of the power at frequency 1.0, whereas PDM correctly identifies the
periodic variation at frequency 1/2.

\begin{figure*}
\centering
\plotone{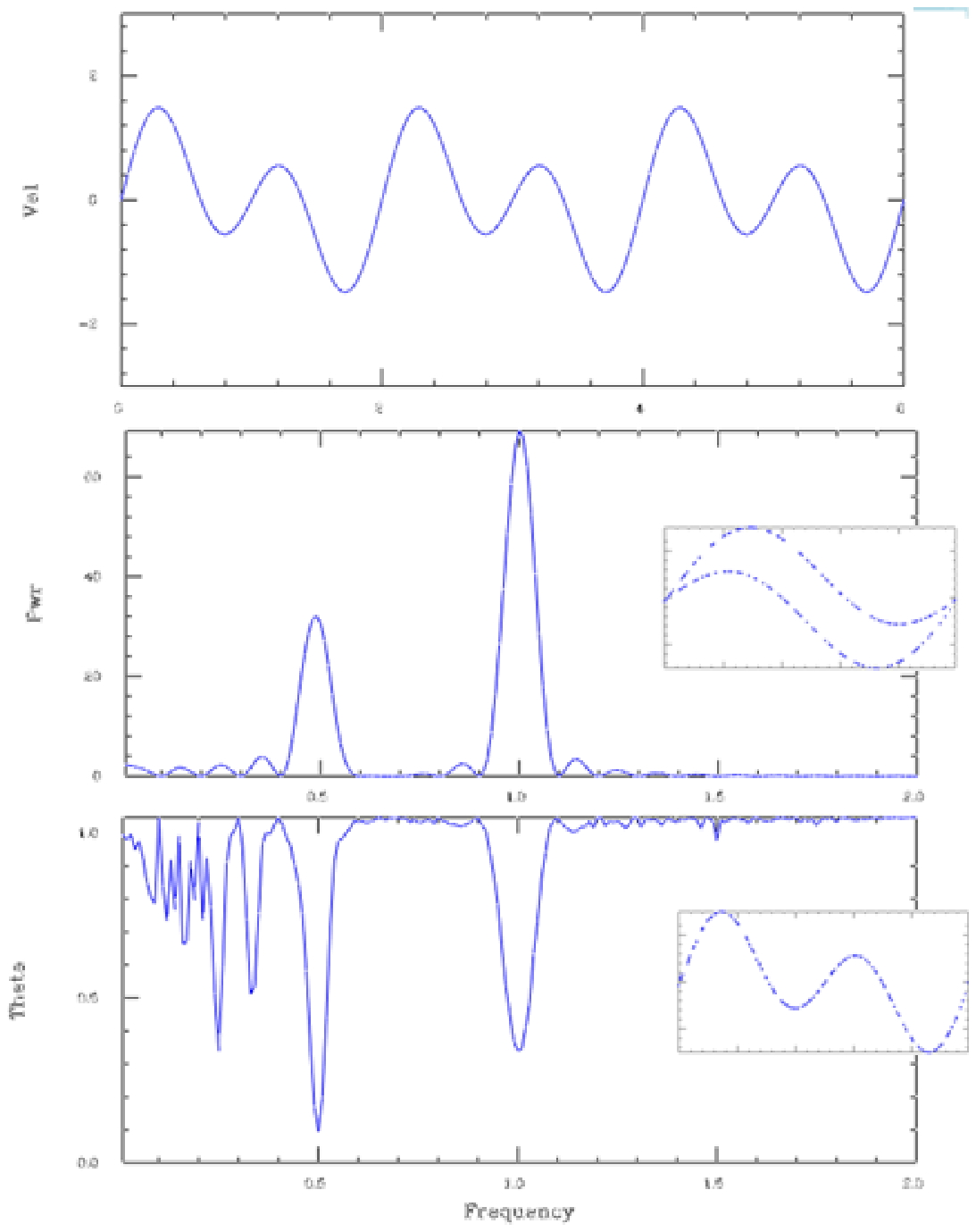}
\vskip0pt
\caption{ Double peaked data variation (top), Lomb analysis (center), PDM result (bottom). 
          The correct frequency is 0.5 (PDM). Inserts show wave form of the main frequency 
          for each case.
%Analysis of a pulse data set: data (top), Lomb power spectrum (center), PDM Theta function (bottom).
}
\label{fig9}
\end{figure*}

\section{Statistical Significance of Results}

It is important to understand the likelihood that a possible periodic variation represents an
actual measured variation, rather than just a low (or high) value in a noise signal containing
no actual variation. To assist in this evaluation, these techniques provide a ``confidence level'' 
or a ``significance'' representing the probability that a pure noise signal would produce the given
result.

To do this, a series of analyses of a pure Gaussian noise signal can be made, and distribution function for various values of the power (Lomb), or of Theta (PDM) can be derived (see Stellingwerf, 1978, Lomb, 1976, Scargle, 1982), and NRC92 Section 13.8). Note that these estimates will be accurate only if the noise observed is pure Gaussian, as assumed in the derivation. In many cases, such as variation due to other modes, Blazhko effect, light contamination, etc. the noise will not be Gaussian, and the analytical estimate is not valid. In many cases, however, a Monte Carlo technique will still be applicable (see below).

As an example, consider the case of 51 data points. The question is whether a signal has (or can) be detected in the presence of noise. So we consider first the case of pure noise. Figure 10 (top) shows the power spectrum from a Lomb analysis of this data set. Note that most of the spectral peaks are value 2 or below, with one peak up to about 4.3. The lower part of Figure 10 shows the PDM analysis of the same data set. Here the Theta variation is above about 0.90.

\begin{figure*}
\centering
\plotone{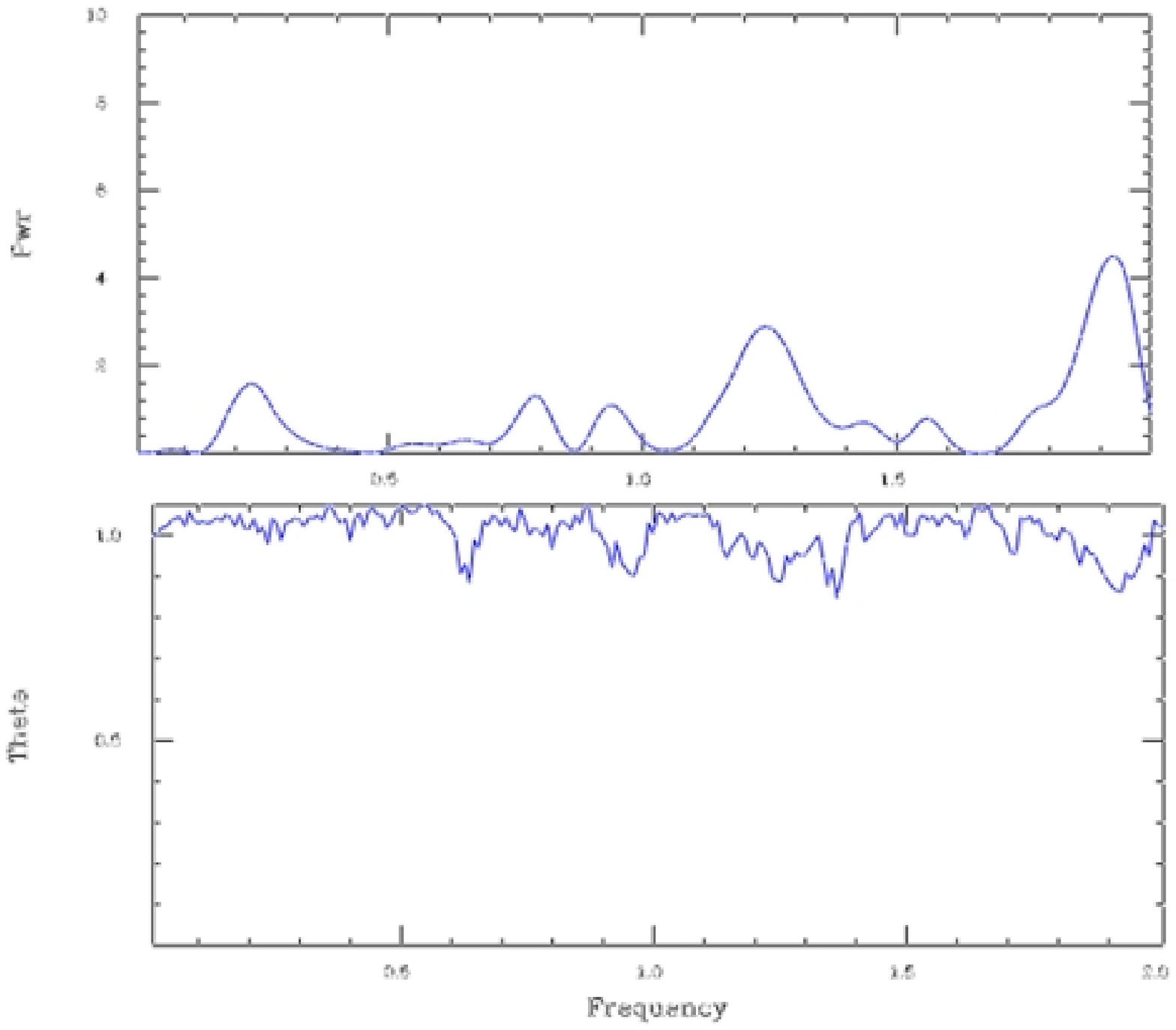}
\vskip0pt
\caption{Top: Lomb power spectrum of 51 points of pure Gaussian noise. Bottom: PDM analysis of the same data set.
}
\label{fig10}
\end{figure*}

The distribution function (probability of a given value) corresponding to the Lomb case is the left curve in Figure 11.
This can be obtained either theoretically (exponential) or directly from the data in Figure 10, and the results agree.
This is not the final result, however, since a correction must be made to account for the number of samples (frequencies),
called the ``Bandwidth Correction'' (see NRC92 for details). This moves the line to the right-hand curve in Figure 11. 
This correction is not small, and there is often some uncertainty in its application. Note that the corrected curve is 
obviously not a simple exponential. We will show below the corrected distribution is an ``extreme value''
distribution, 
(e.g. Weibull distribution). Significant results must fall to the right of the tail of the distribution (value greater 
than about 7.5). Note that this value is significantly larger than the computed maximum of about 4.5.

\begin{figure*}
\centering
\includegraphics[width=14.5cm]{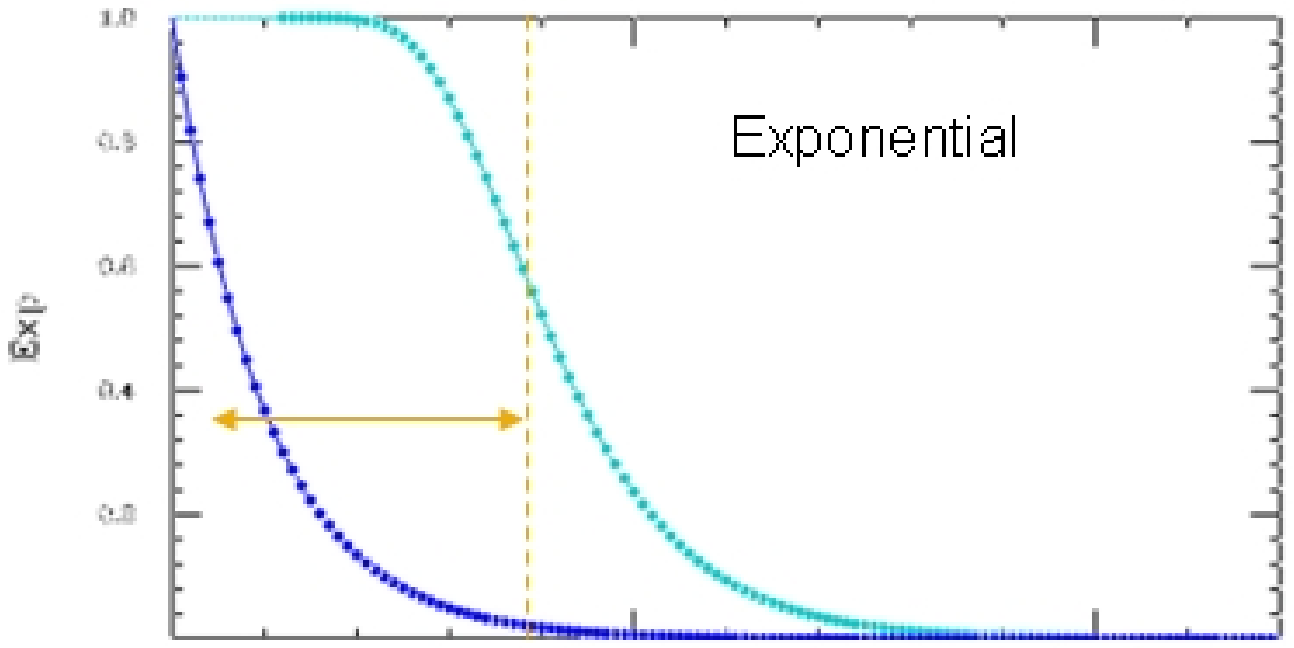}
%\plotone{newfig11.eps}
\vskip0pt
\caption{Distribution (left) of the power spectrum shown in Figure 10. Right: distribution with
         bandwidth correction. Orange arrow = range of data.
%Top: Distribution (left) of the power spectrum shown in Figure 10. Right: distribution with bandwidth correction. Orange arrow = range of data.
}
\label{fig11}
\end{figure*}

The PDM distribution is the Beta Distribution shown in Figure 12 (right curve). The left-hand curve
is the distribution after the bandwidth correction is applied. ``Significant'' results will now
fall to the left of the tail, values of less than about 0.65. Again, this is much smaller than the
computed minimum of about 0.90. Note that this distribution falls off more rapidly than the exponential
in the periodogram case. This should result in a cleaner distinction between real and random results.

\begin{figure*}
\centering
\includegraphics[width=14.5cm]{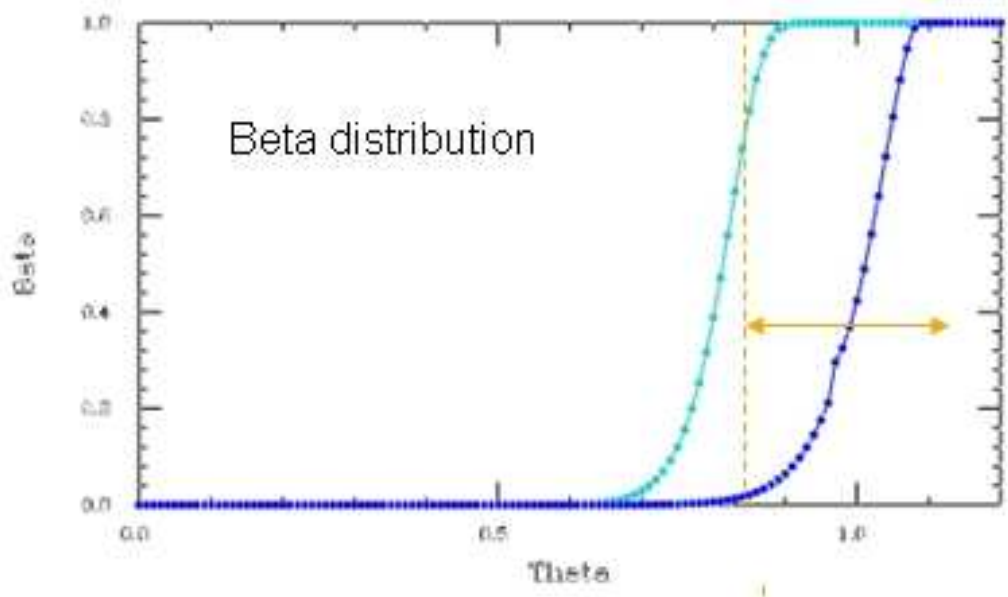}
%\plotone{fig12.eps}
\vskip0pt
\caption{Distribution (right) of  PDM Theta for the 51 point noise case. Left: distribution with bandwidth correction. Orange arrow = range of data.
}
\label{fig12}
\end{figure*}

As mentioned above, we can imagine a variety of cases in which the exponential or Beta function
analyses are not expected to be accurate. In these cases the Monte Carlo approach developed by
Nemec \& Nemec (1985) can be used. In this method, multiple analyses of the data are done, with
phase redistribution applied to each run to obtain independent random samples. For each case the
minimum value of Theta is noted. The distribution of these minima is then used to compute the
probability that an observed minimum is real. This is an ``extreme value'' distribution. The 
result of applying this approach to a PDM analysis of the 51 point Gaussian noise test case
(500 passes) is shown in Figure 13 (left hand curve. Note that this curve corresponds closely to
the analytic Beta distribution result shown in Figure 12 {\em after the bandwidth correction.} For
comparison, a distribution curve was computed for a single PDM run, and this is the right hand
curve in Figure 13. It corresponds closely with the Beta result in Figure 12 {\em before correcting.}
This provides another interpretation of the bandwidth correction -- it is the extreme value
distribution corresponding to the exponential or Beta distribution of the analysis. This explanation
is less ambiguous and clearer than the usual interpretation. For a quick estimate, the analytic
distributions usually suffice, but for important borderline cases, the Monte Carlo approach is
favored. In PDM2 250 MC passes is the default, with an additional check with 500 passes recommended
to validate important cases.

\begin{figure*}
\centering
\plotone{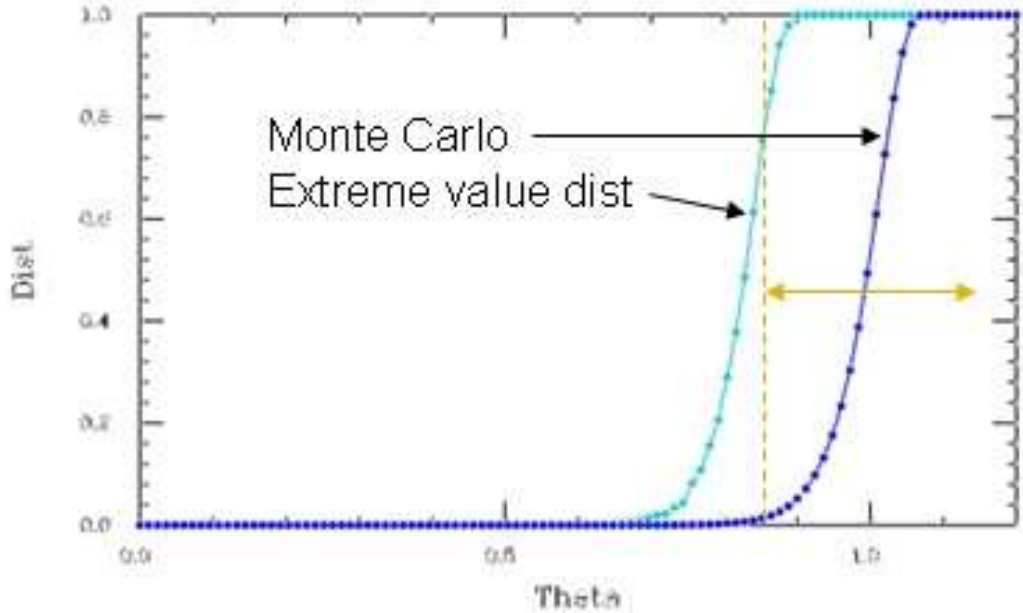}
\vskip0pt
\caption{Distribution (left) of Monte-Carlo analysis of the PDM analysis for the 51 point noise case. . Right: distribution of a single PDM run.  Orange arrow = range of data.}
\label{fig13}
\end{figure*}

\vbox{
\vskip0.35cm
}
\section{Tests on IC4499 Data / Period Changes}

Several test cases on data kindly supplied by the authors of Kunder, et.al. (2011) were 
analyzed and will be briefly shown here. Three variables were considered from the IC4499 
data set: V04, V48, and V83. V04 is a type a RR Lyrae with a well determined period and 
period change. The other two are Blazhko variables with known periods, but no period change
estimates.

Figure 14 show the data and PDM analysis for V04. The data set is shown in the top panel, 
colored on the sigma (uncertainty) for each point. The sigmas are small for this data set, 
but were taken into account in the PDM analysis. The middle panel shows the result of a 
broad-range frequency scan with nine data segments to broaden the features. The bottom panel
shows the result of a fine scan near the candidate frequency, resulting in a final estimate
of 1.60355 / day. The published value is 1.60354.

\begin{figure*}[ht]
\centering
\includegraphics[width=15.5cm]{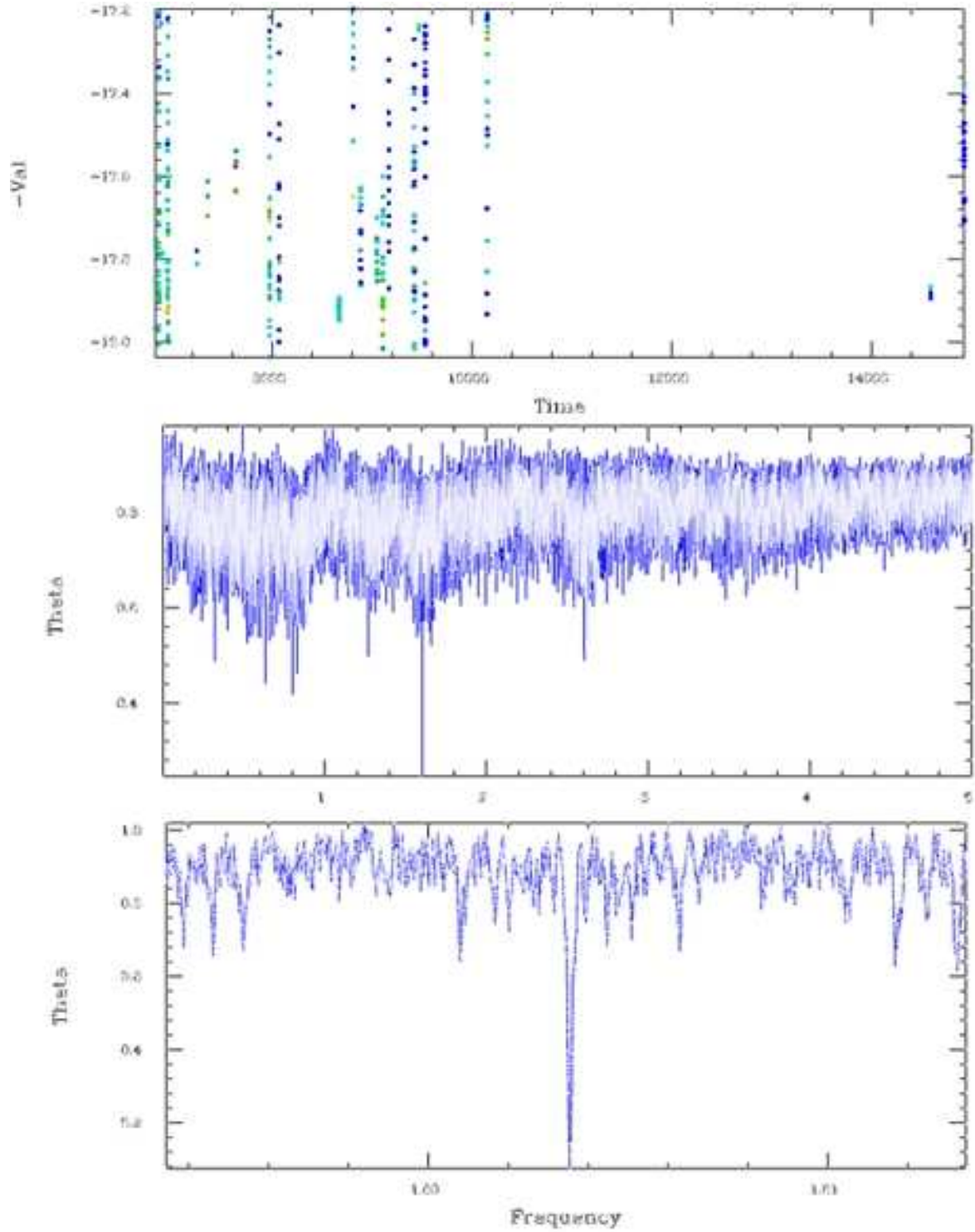}
%\plotone{fig14.eps}
\vskip0pt
\caption{Top: Data set for v04, Middle: Theta variation for the segmented period search. 
       Bottom: Theta variation for the full data set period scan.}
\label{fig14}
\end{figure*}

In the 2011 paper an O-C analysis is used to estimate the period change. If the period is 
given by $P = P_0 + \beta$ t, the value for $\beta$ was found to be 0.18 +/- .05 d/My. PDM2
has a new option to vary $\beta$ and see if the variance of the fit is reduced. The result 
of this scan is shown in Figure 15. A clearly defined minimum is seen around Beta = 1.5 or 
so with an uncertainty of about +/- 0.10. This is less accurate than the O-C analysis result,
but is consistent with it.

\begin{figure*}
\centering
\includegraphics[width=13.4cm]{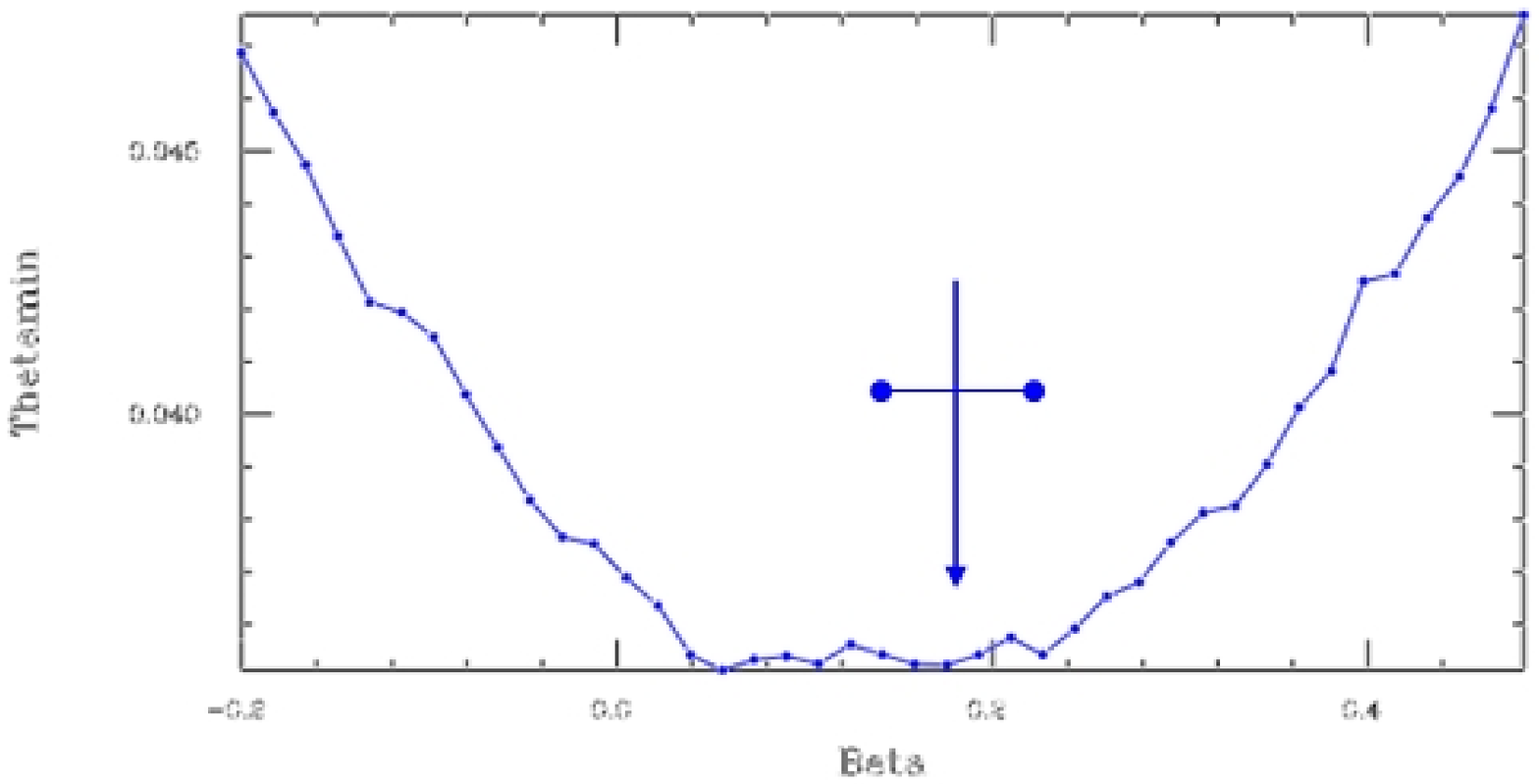}
%\plotone{fig15.eps}
\vskip0pt
\caption{Theta-min versus Beta / period change analysis for v04. Arrow and bar indicates the O-C result.}
\label{fig15}
\end{figure*}

%\vbox{
%\vskip0.5cm
%}
%\clearpage

Figure 16 shows the resulting light curve with the period change taken into account. The 
data are very tight, and the scatter is reduced from the constant period case.

\begin{figure*}
\centering
\includegraphics[width=13.4cm]{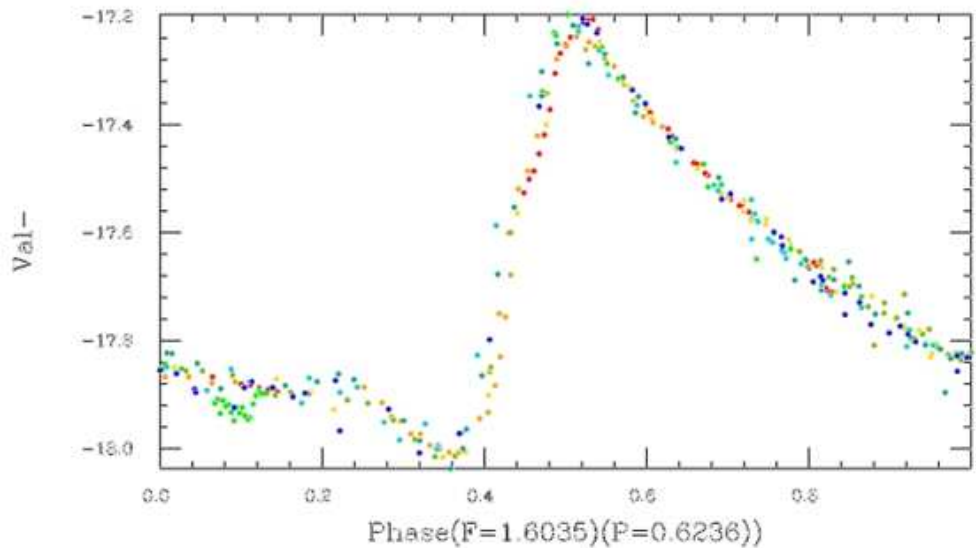}
%\plotone{fig16.eps}
\vskip0pt
\caption{ Final phase plot of v04 data, with period change in effect.}
\label{fig16}
\end{figure*}

Variable v48 is a Blazhko variable and the O-C analysis could not produce an estimate for 
the period change rate. The PDM analysis is similar to that of v04, but the mean light curve,
as shown in Figure 17 shows a typical Blazhko variation in amplitude. The point colors are 
time values, so the red low amplitude peak represents late points, so a decreasing period 
may be suspected.

\begin{figure*}
\centering
\includegraphics[width=12cm]{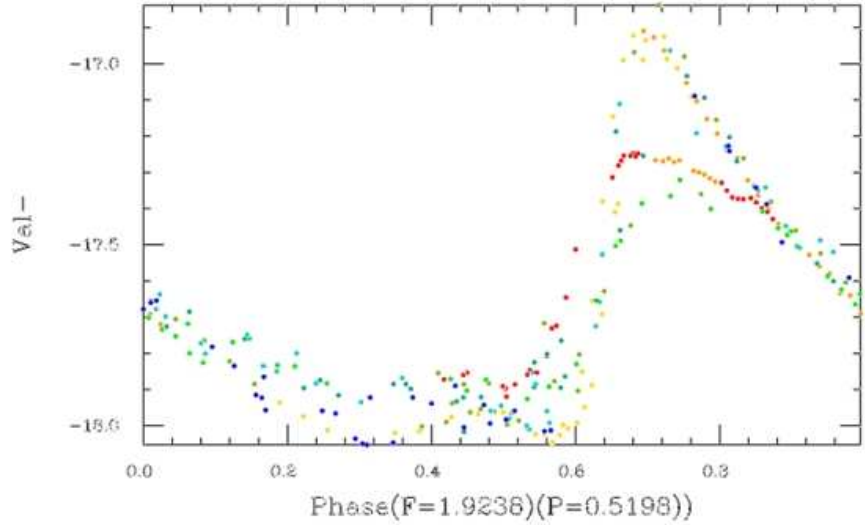}
%\plotone{fig17.eps}
\vskip0pt
\caption{Light curve for v48.}
\label{fig17}
\end{figure*}

A period change scan was run on this data, with the results shown in Figure 18.  By looking
at the scatter at all phases simultaneously, PDM is able to determine the period change to
be Beta = -0.13+/-0.05 with a well-defined minimum. With this period change in effect, the 
scatter on the rising branch is eliminated.

\begin{figure*}
\centering
\includegraphics[width=12cm]{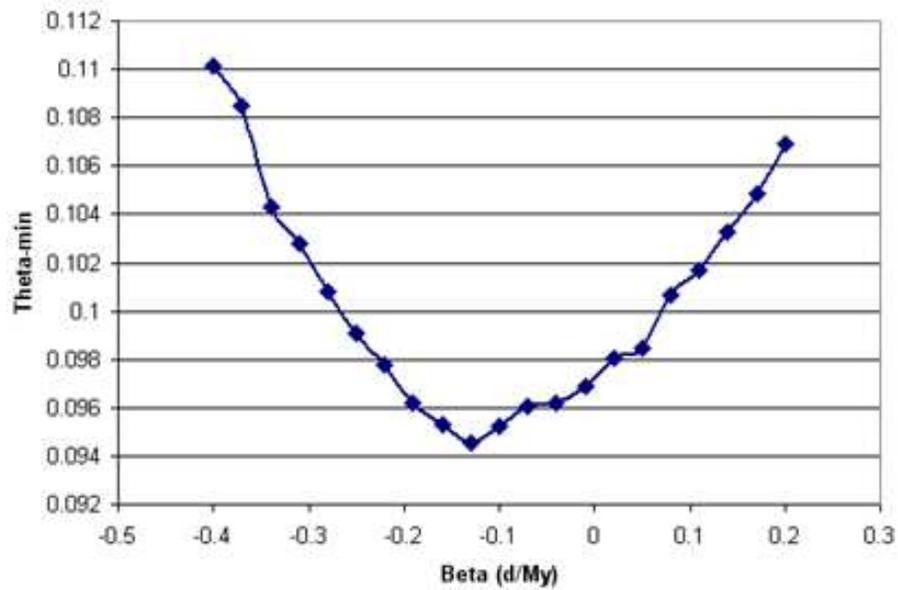}
%\plotone{newfig18.eps}
\vskip0pt
\caption{ Theta-min versus Beta for v48 -- period change scan.}
\label{fig18}
\end{figure*}

Variable 83 is also classified as Blazhko. The light curve resulting from a PDM analysis is shown in Figure~19. Here the
scatter is mainly at minimum light, but the red points are again found to the left of the other data, suggesting a
decreasing period. This is confirmed by the period scan shown in Figure~20, producing a result of Beta = -0.20 +/- 0.10. 

\begin{figure*}
\centering
\includegraphics[width=14.0cm]{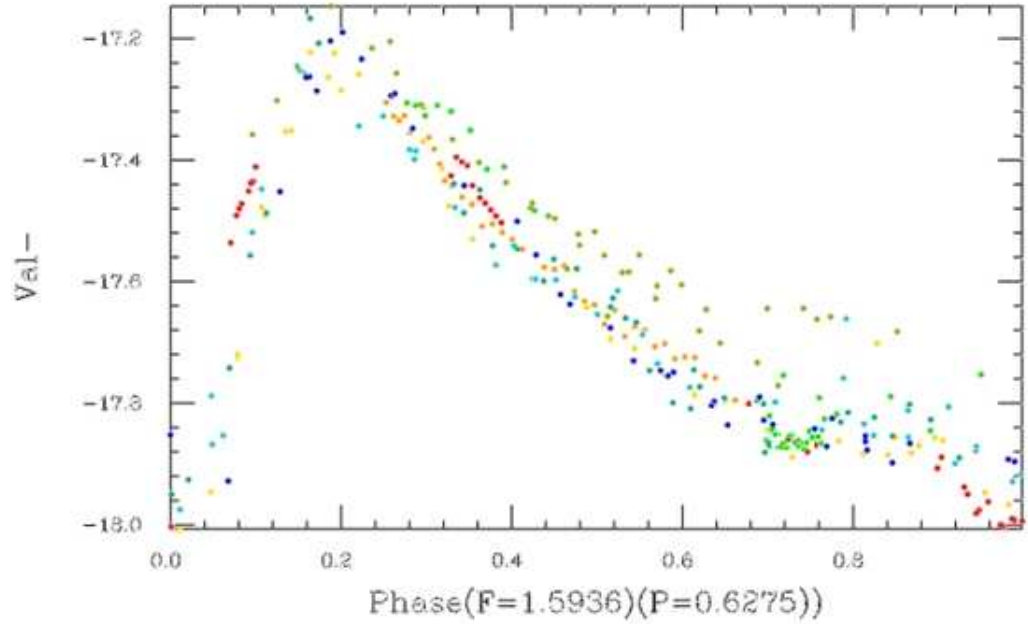}
%\plotone{fig19.eps}
\vskip0pt
\caption{Light curve for v83.}
\label{fig19}
\end{figure*}

\begin{figure*}[hb]
\centering
\includegraphics[width=14.0cm]{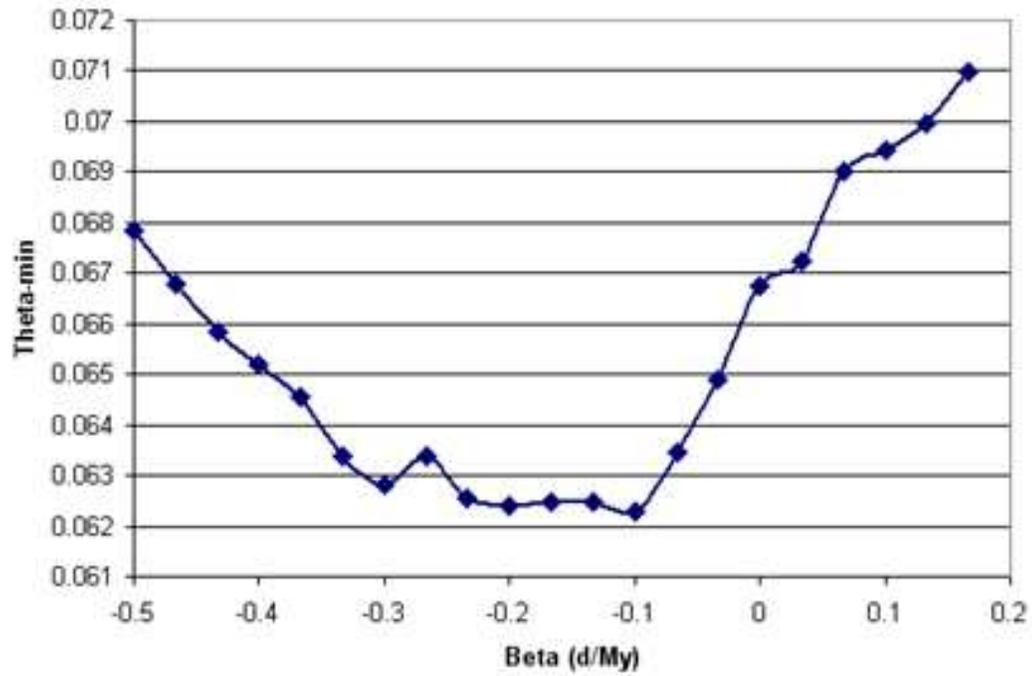}
%\plotone{fig20.eps}
\vskip0pt
\caption{Period change analysis for v83.}
\label{fig20}
\end{figure*}

\vfill\eject

\section{Conclusions}

A new version of the period analysis program PDM has been developed and is available as a Windows executable or C source code for a Unix compile. The tests discussed here show that this approach is preferable in many cases for a straight period search, while the Lomb or Fourier techniques are preferable for cases with broad frequency spreads whose power distributions are desired.

The new version of PDM has a number of new features and updates several known problems with the original version of the technique. We show, in particular that it is applicable to determining period changes for Blazhko variables that present difficult cases for fixed-phase analyses.

\end{document}